\documentclass[paper=a4,11pt,twoside]{article} 

\usepackage[utf8]{inputenc}

\usepackage[dutch,british]{babel} 

\usepackage[top=2.54cm,bottom=2.54cm,left=3.17cm,right=3.17cm]{geometry} 

\usepackage{hyperref} 

\usepackage{fancyhdr} 

\usepackage{emptypage} 

\usepackage{enumitem}
\usepackage{caption}

\usepackage[fleqn]{amsmath}
\usepackage{amssymb}
\usepackage{amsfonts}
\usepackage{mathtools}
\usepackage{array}
\usepackage{siunitx}
\usepackage{algorithmic}
\usepackage{float}

\usepackage{setspace}

\usepackage{graphicx}
\usepackage{graphics}
\usepackage{subfigure}

\usepackage{booktabs} 
\usepackage{multirow}
\usepackage{longtable} 

\usepackage[square,comma,numbers,sort&compress]{natbib}

\usepackage{xcolor}

\usepackage[noblocks]{authblk}

\usepackage[toc,page]{appendix}

\usepackage{ulem}

\numberwithin{table}{section} 
\numberwithin{equation}{section} 
\graphicspath{{../Figs/}} 
\DeclareGraphicsExtensions{.eps,.mps,.pdf,.jpg,.png} 
\definecolor{IgorGreen}{RGB}{76,153,0}
\definecolor{IgorBlue}{RGB}{0,0,204}
\definecolor{IgorGreen1}{RGB}{0,153,77}
\definecolor{IgorPurple}{RGB}{103,0,204}

\begin{document}
\title{Non-consensus opinion model on directed networks}

\author[1]{Bo Qu}      
\author[2]{Qian Li}            
\author[3]{Shlomo Havlin}
\author[2]{H. Eugene Stanley} 
\author[1,2]{Huijuan Wang}
\date{}
\affil[1]{Delft University of Technology, Delft, Netherlands}
\affil[2]{Department of Physics and Center for Polymer Studies, Boston
  University, Boston, Massachusetts 02215, USA}                 
\affil[3]{Department of Physics, Bar Ilan University, Ramat Gan, Israel}

\date{4 April 2014}

\maketitle

\begin{abstract}

Dynamic social opinion models have been widely studied on undirected
networks, and most of them are based on spin interaction models that
produce a consensus. In reality, however, many networks such as Twitter
and the World Wide Web are directed and are composed of both
unidirectional and bidirectional links. Moreover, from choosing a coffee
brand to deciding who to vote for in an election, two or more competing
opinions often coexist. In response to this ubiquity of directed
networks and the coexistence of two or more opinions in decision-making
situations, we study a non-consensus opinion model introduced by Shao et
al. \cite{shao2009dynamic} on directed networks. We define
directionality $\xi$ as the percentage of unidirectional links in a
network, and we use the linear correlation coefficient $\rho$ between
the indegree and outdegree of a node to quantify the relation between
the indegree and outdegree. We introduce two degree-preserving rewiring
approaches which allow us to construct directed networks that can have a
broad range of possible combinations of directionality $\xi$ and linear
correlation coefficient $\rho$ and to study how $\xi$ and $\rho$
impact opinion competitions. We find that, as the directionality $\xi$
or the indegree and outdegree correlation $\rho$ increases, the majority
opinion becomes more dominant and the minority opinion's ability to
survive is lowered.

\end{abstract}

\section{Introduction}

Network theory based on graph theory uses a graph to represent symmetric
or asymmetric relations between objects shown by undirected and directed
links, respectively. The study of social networks is one of the most
important applications of graph theory.  Social scientists began
refining the empirical study of networks in the 1970s, and many of the
mathematical and physical tools currently used in network science were
originally developed by them \cite{newman2009networks}. Social network
science has been used to understand the diffusion of innovations, news,
and rumors as well as the spread of disease and health-related human
behavior
\cite{givan2011predicting,kitsak2010identification,cohen2003efficient,braunstein2003optimal,van2011n,gross2008adaptive,wang2013effect}.
The decades-old hot topic of opinion dynamics continues to be a central
focus among researchers attempting to understand the opinion formation
process.  Although it may seem that treating opinion as a variable or a
set of variables is too reductive and the complexity of human behavior
makes such an approach inappropriate, often human decisions are in response
to limited options: to buy or not to buy, to choose Windows or Linux, to
buy Procter \& Gamble or Unilever, to vote for the Republican or the
Democrat.

Treating opinion as a variable allows us to model patterns of opinion
formation as a dynamic process on a complex network with nodes as agents
and links as interactions between agents. Although the behavior dynamics
of human opinion are complex, statistical physics can be used to
describe the ``opinion states'' within a population and also the
underlying processes that control any transitions between them
\cite{Castellano2009,boccaletti2006complex,pastor2001epidemic,albert2002statistical,galam2005local,dorogovtsev2003evolution,borghesi2012between}.
Over the past decade numerous opinion models have combined complex
network theory and statistical physics. Example include the Sznajd model
\cite{sznajd2000opinion}, the voter model
\cite{liggett1999stochastic,lambiotte2008dynamics,schweitzer2009nonlinear}, the majority rule
model \cite{galam2002minority,krapivsky2003dynamics}, the social impact
model \cite{latane1981psychology,nowak1990private}, and the bounded
confidence model \cite{deffuant2000mixing,hegselmann2002opinion}.  All
of these models ultimately produce a consensus state in which all agents
share the same opinion. In most real-world scenarios, however, the final
result is not consensus but the coexistence of at least two differing
opinions.

Shao et al. \cite{shao2009dynamic} proposed a non-consensus opinion (NCO)
model that achieves a steady state in which two opinions can
coexist. Their model reveals that when the initial population of a
minority opinion is above a certain critical threshold, a large
steady-state spanning cluster with a size proportional to the total
population is formed \cite{shao2009dynamic}. This NCO complex network
model belongs to the same universality class as percolation
\cite{shao2009dynamic,bunde1991fractals,stauffer1994introduction}, and
it and its variants, have received much attention. Among the variants
are a NCO model with inflexible contrarians \cite{li2011strategy} and
a NCO model on coupled networks \cite{Li2013,ben2011exact}. 

To date the model has not been applied on directed
networks. Directed networks are important because many real-world
networks, e.g., Twitter, Facebook, and email networks, are directed
\cite{PhysRevE.88.062802}.  In contrast to undirected networks, directed
networks contain unidirectional links. In opinion models, a
unidirectional link between two nodes indicates that the influence
passing between the two nodes is one-way. A real-world example might be
a popular singer who influences the opinions the fans hold, but the fans
do not influence the singer's opinion. In contrast, bidirectional links
occur when the influence between two agents is both ways.  Real-world
unidirectional links are ubiquitous and strongly influence opinion
formation, i.e., widespread one-way influence has a powerful effect on
opinion dynamics within a society.

Our goal here is to examine how the NCO model behaves on directed
networks. We compare the results of different networks in which we vary
the proportion of unidirectional links. We also measure the influence of
asymmetry between indegree and outdegree. We find that when the indegree and outdegree of
each node are the same, an increase in
the number of unidirectional links helps the majority opinion spread and when the fraction of unidirectional links is at a certain
level, increasing the asymmetry between indegree and outdegree increases the
minority opinion's ability to survive. We also observe that changing the
proportion of the unidirectional links or the relationship between the
indegree and outdegree of the nodes causes phase transitions. 

\section{Basic definitions and notations}
\subsection{The NCO model}

In a NCO model \cite{shao2009dynamic} on a single network with $N$ nodes, each with binary
opinions, a fraction $f$ of nodes has opinion $\sigma_{+}$ and a
fraction $1-f$ has opinion $\sigma_{-}$. The opinions are initially
randomly assigned to each node. At each time step, each node adopts the
majority opinion, when considering both its own opinion and the opinions of 
its nearest neighbors (the agent's friends). A node's opinion does not
change if there is a tie. Following this opinion formation rule, at each
time step the opinion of each node is updated. The updates occur
simultaneously and in parallel until a steady state is reached. Note
that when the initial fraction $f$ is above a critical threshold,
$f\equiv f_{c}$ (even minority), both opinions continue to exist in 
the final steady state.

\begin{figure}[htbp]
\centering
\includegraphics[scale=.8]{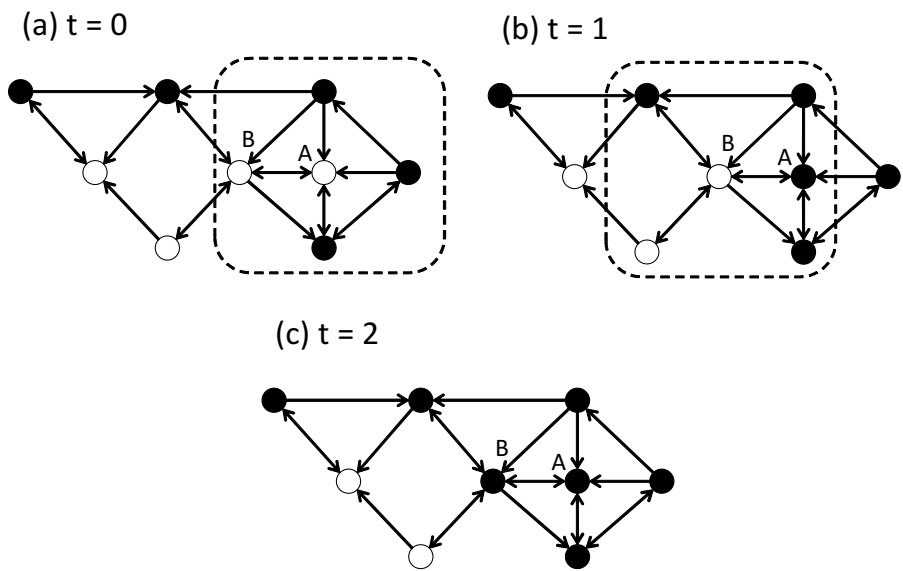}
\caption{Schematic plot of the dynamics of the NCO model on a directed graph with 9 nodes.}
\label{fig:NCO-1}
\end{figure}

Figure~\ref{fig:NCO-1} shows an example of the dynamic process of the
NCO model on a small directed network with nine nodes. Here we consider
the in-neighbors of a node as the friends influencing the node, and the
out-neighbors as the friends influenced by the node. At time $t=0$ four
nodes are randomly assigned the opinion $\sigma_{+}$ (empty circle), 
and the other five nodes the opinion $\sigma_{-}$. At time
$t=0$ node A has opinion $\sigma_{+}$ but is in a local minority and
thus updating it means changing its opinion to $\sigma_{-}$.  At time
$t=1$ node B belongs to a local minority and thus needs updating. At
time $t=3$ all nodes hold the same opinion as their local majority, and
the system has reached a final non-consensus steady state.

\subsection{The directionality $\xi$ and indegree outdegree correlation $\rho$ }

To quantitatively measure the one-way influence in a network, we define
the directionality $\xi$ as the ratio between unidirectional links and
all links. The directionality is $\xi=L_{\rm unidirectional}/L$, where the
normalization $L=L_{\rm unidirectional}+2L_{\rm bidirectional}$, because
a bidirectional link can be considered as two unidirectional links. Because
we want to determine how much one-way influence affects the NCO model,
we consider as a variable the fraction of one-way links $\xi$, where $\xi=0$
represents undirected networks. Although the
sum of indegree and the sum of outdegree are equal in a directed
network, the indegree and outdegree of a single node are usually not the same.  To quantify the possible difference between the node's indegree and
outdegree, we use the linear correlation coefficient $\rho$ between
them,
\begin{equation}
\rho = \frac{\sum_{i=1}^{N}(k_{i,{\rm in}}-\langle k\rangle)(k_{i,{\rm out}}-\langle k\rangle)}{\sqrt{\sum_{i=1}^{N}(k_{i,{\rm in}}-\langle k\rangle)^2}\sqrt{\sum_{i=1}^{N}(k_{i,{\rm out}}-\langle k
\rangle)^2}}      
\end{equation}
where $k_{i,{\rm in}}$ and $k_{i,{\rm out}}$ are the indegree and
outdegree of node $i$ respectively. The average degree $\langle k
\rangle$ is the same for both indegree and outdegree. Note that when
$\rho=1$ the indegree is linearly dependent on the outdegree for all
nodes, and when $\rho=0$ the indegree and outdegree are independent of
each other. In this paper we confine ourselves to the case in which the
indegree and outdegree follow the same distribution. In this case,
$\rho=1$ implies that $k_{i,{\rm in}}=k_{i,{\rm out}}$ holds for every
node $i$.
  
\section{Algorithm Description}

Inspired by earlier research on directed networks
\cite{newman2001random,PhysRevE.88.062802,sanchez2002nonequilibrium,schwartz2002percolation,PhysRevE.85.046107,Li2013},
we propose two algorithms to construct directed networks. One is a
rewiring algorithm that can be applied to any existing undirected
network to obtain a directed network with any given directionality but each node has the same indegree and outdegree as the original undirected
network. The other constructs directed networks with a given
directionality and indegree-outdegree correlation, and with the same
given indegree and outdegree distribution. Note that all networks
considered in this paper contain neither self-loops nor multiple links
in one direction between two nodes.

\subsection{Directionality-increasing rewiring (DIR)} 
\label{sec:IncreaseDir}

Here we introduce a rewiring approach that changes the directionality
but does not change the indegree and outdegree of any node. It was first
proposed in Ref.~\cite{van2010influence}, and also employed by
Ref.~\cite{PhysRevE.88.062802}. Here we improve it to gradually increase
the directionality, via a technique we call directionality-increasing
rewiring (DIR).

\begin{figure}[htbp]
\centering
\includegraphics[scale=.8]{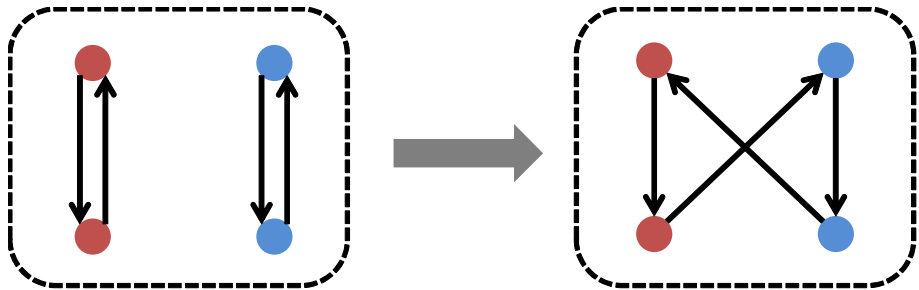}
\caption{(Color online) Directionality-increasing rewiring (DIR)}
\label{fig:NCO-2}
\end{figure}

Many undirected network models with various properties have been
designed. Examples include the Erd\"os-R\'enyi model
\cite{erdds1959random}, the B\'arabasi-Albert scale-free model
\cite{barabasi1999emergence}, and the small-world model
\cite{watts1998collective}. If the links of an undirected graph are
considered bidirectional, for an arbitrary undirected graph the indegree
and outdegree correlation will be $\rho=1$. Figure~\ref{fig:NCO-2} demonstrates
an approach that changes the directionality but does not change the
indegree and outdegree of any node nor $\rho$. We randomly choose two
bidirectional links connecting four nodes and treat them as four
unidirectional links. Note that this differs from the technique
presented in Ref.~\cite{PhysRevE.88.062802} in that we choose two
bidirectional links instead of two random links that may also contain
unidirectional links so that the directionality increases after each
step. Then we choose two unidirectional links, one from each
bidirectional link, and rewire them as follows. For both unidirectional
links the head of one link is replaced with the head of the other. If
this rewiring introduces multiple links in any direction between any two
nodes, we discard it and randomly choose two other bidirectional
links. We can increase the number of unidirectional links by repeating
the rewiring step and increasing the directionality in each step. The
directionality $\xi$ can be varied from 0 to 1.  In general, DIR can be
applied to any directed network to further increase its directionality.

\subsection{Constructing an asymmetric indegree and outdegree network
  and rewiring it to decrease its directionality (ANC-DDR)}
\label{sec:DecreaseDir}

We have shown how to obtain a desired directionality $\xi$ when the
indegree and outdegree correlation is $\rho=1$. We further propose an
algorithm to construct a network with a given combination of $\xi$ and
$\rho$, where $\rho\neq 1$. Inspired by the work presented in
Ref.~\cite{schwartz2002percolation}, which focuses on generating
directed scale free (SF) networks with correlated indegree and outdegree
sequences, we extend it to a scenario in which the indegree and
outdegree sequences follow a distribution that is arbitrary but the
same, and we control not just the correlation between the indegree and
outdegree but also the directionality, which was ignored in
Ref.~\cite{schwartz2002percolation}. We generate an indegree sequence
(following a Poisson distribution or power law) and a null outdegree
sequence. We then copy a fraction $\rho$ of the indegree sequence to the
outdegree sequence, and shuffle the fraction $1-\rho$ of the indegree
sequence as the rest of the outdegree sequence. We thus create an
outdegree sequence, a fraction $\rho$ of which is identical to the
corresponding part of the indegree sequence and a fraction $1-\rho$ of
which is independent of the indegree sequence. After randomly connecting
all nodes (given their indegree and outdegree), as in the configuration
model \cite{newman2001random}, we obtain a network with a
directionality\footnote{$E(\xi)=1-\langle k\rangle^{2}N/(N-1)^{2},
  \lim_{N\to +\infty}E(\xi)=1$.} $\xi\approx 1$ and an indegree and
outdegree correlation close to $\rho$. Note that we can further control
the indegree and outdegree correlation in a small range close to $\rho$
by discarding networks with indegree and outdegree correlations outside
the expected range. This enables us to construct a network with the
indegree and outdegree correlation $\rho$ ($0\lessapprox\rho\leq1$), a
technique we call asymmetric indegree-outdegree network constructing
(ANC).

\begin{figure}[htbp]
\centering
\subfigure[]{
\label{fig:NCO-3_a}
\includegraphics[scale=.4]{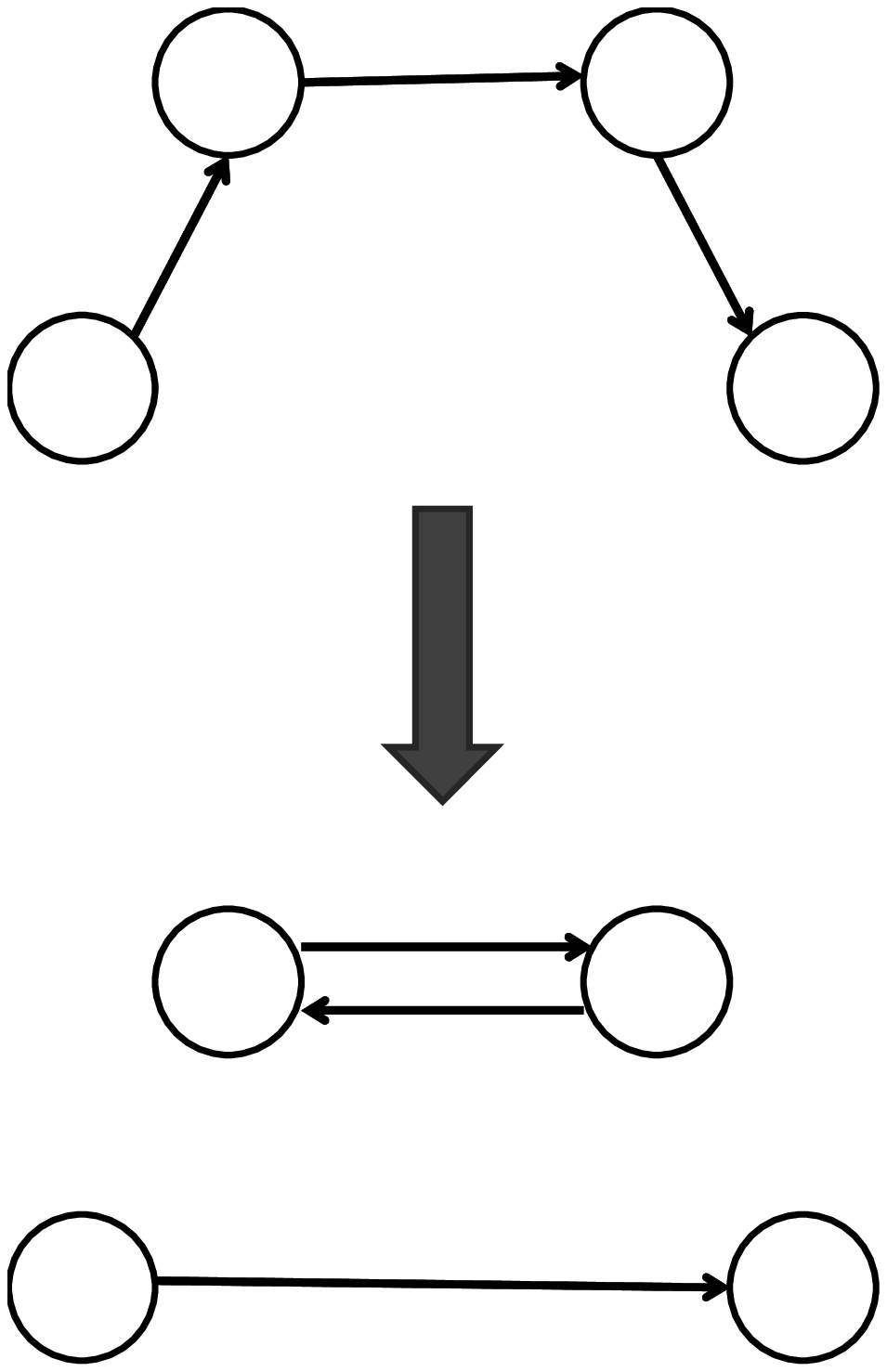}
}
\hspace{80pt}
\subfigure[]{
\label{fig:NCO-3_b}
\includegraphics[scale=.3]{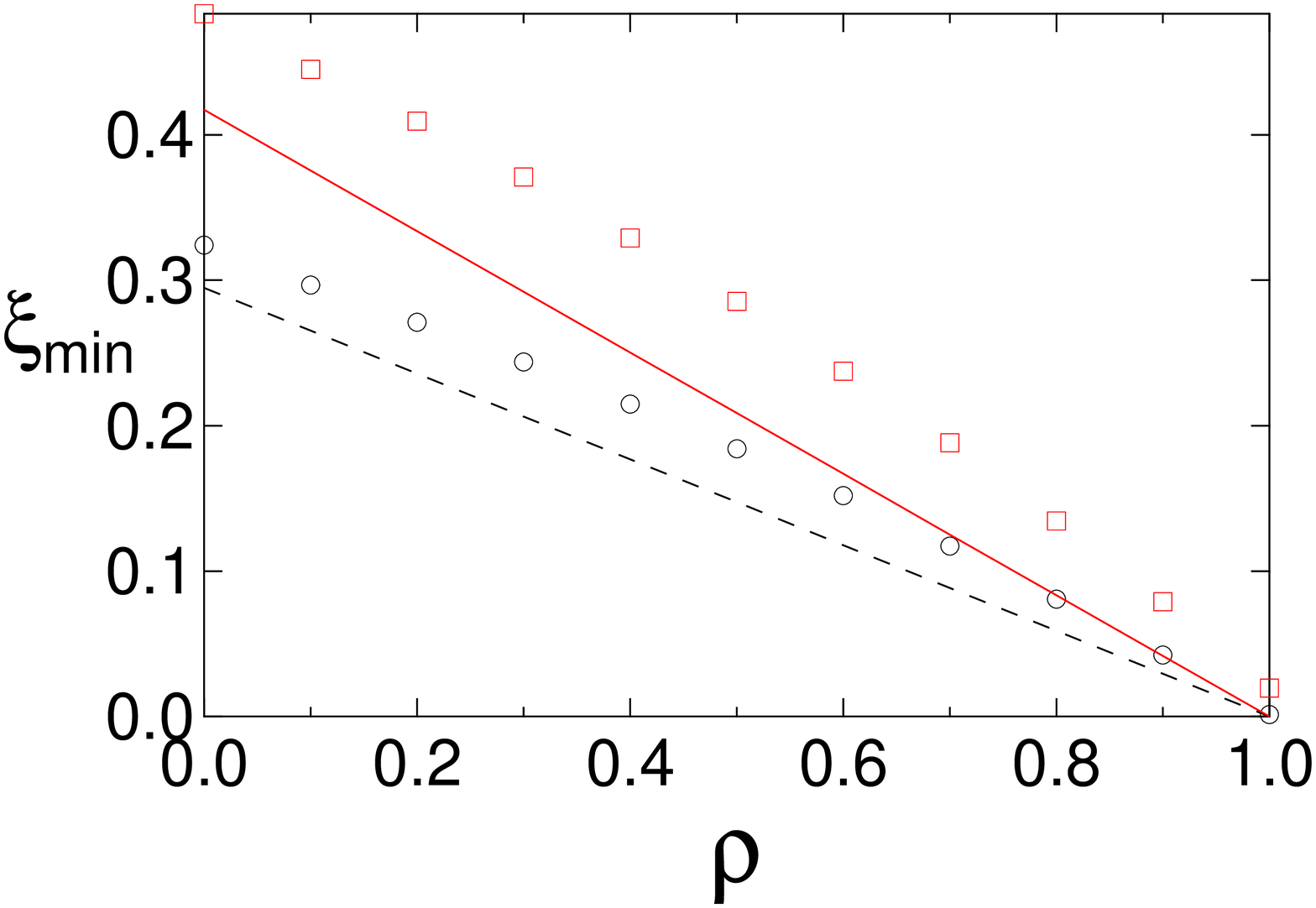}
}
\caption{(Color online) (a) The degree-preserving rewiring for
  decreasing the directionality. (b) Plot of the minimal directionality
  $\xi_{\rm min}$ obtained by simulating ANC-DDR, for binomial networks
  ($\circ$, $\langle k\rangle=4$, $10^5$ nodes) and SF networks
  ($\textcolor{red}{\square}$, $\lambda = 2.63$, $10^5$ nodes) with 100
  realizations, and the theoretical minimum possible directionality
  $\xi_{\rm min}$, Equ. (\ref{equ:31}), for binomial networks (the solid
  line) and for SF networks (the dash line) as a function of the
  indegree and outdegree correlation $\rho$. }
\end{figure}

We use the following rewiring steps to further tune the directionality
without changing the indegree and outdegree of each node or the indegree
and outdegree correlation $\rho$. The goal is to decrease the
directionality by repeatedly rewiring two unidirectional links into one
bidirectional link. In each step, we choose four nodes linked with at
least three directed links as shown on the top half of
Fig.~\ref{fig:NCO-3_a}. We rewire these three links to the positions
shown at the bottom of Fig.~\ref{fig:NCO-3_a}. If this rewiring
introduces multiple links between any two nodes in any direction we
discard the rewiring, select four new nodes, and repeat the
step\footnote{An efficient rewiring program is available upon
  request}. This rewiring produces at least one more bidirectional link
and thus decreases the directionality. We call this procedure
directionality-decreasing rewiring (DDR). We combine DDR with ANC and
call the entire algorithm ANC-DDR.

Using ANC we can construct a network with a specified indegree and
outdegree correlation $\rho$, where the indegree and the outdegree
follow the same given distribution and, using DDR, we can change the
directionality $\xi$ in a range dependent on the given $\rho$ without
changing the indegree and outdegree. The range within which we can tune
$\xi\in[\xi_{\rm min}, 1]$ depends on the given $\rho$. For example, for
binomial networks\footnote{\textit{Binomial networks\/} are directed
  networks with the same Poissonian indegree and outdegree
  distributions.}, $\xi$ can be changed from $0$ to $1$ when $\rho=1$,
but the minimum value of $\xi$ must be approximately $0.3$ and any
smaller $\xi$ value is disallowed when $\rho=0$. We explore the relation
between the minimal possible directionality $\xi$ and a given indegree
and outdegree correlation $\rho$ first via numerical
simulations\footnote{In each realization of the simulations, we apply
  DDR repeatedly on the network constructed by ANC until the four-node
  structure in Fig.~\ref{fig:NCO-3_a} cannot be found after a number $M$
  of consecutive attempts, then the directionality $\xi$ is considered
  the minimal directionality $\xi_{\rm min}$ corresponding to the given
  $\rho$. For each given $\rho$, we perform $100$ realizations and
  calculate the average of the minimal directionality $\xi_{\rm min}$.}
in both binomial and SF networks\footnote{\textit{SF networks\/} are
  directed networks whose indegree and outdegree distributions follow
  the same power law.}. Figure~\ref{fig:NCO-3_b} shows the linear
relationship in both types of network. Binomial networks are
characterized by a Poisson degree distribution with $P(k)=e^{-\langle k
  \rangle}\langle k \rangle ^{k}/k!$, where $k$ is the node degree and
$\langle k \rangle$ is the average degree. The degree distribution of SF
networks is given by $P(k)\scriptsize{\sim}k^{-\lambda}, k\in[k_{\rm
    min},k_{\rm max}]$, where $k_{\rm min}$ is the smallest degree,
$k_{\rm max}$ is the degree cutoff, and $\lambda$ is the exponent
characterizing the broadness of the distribution
\cite{barabasi1999emergence}. In this paper we use the natural cutoff at
approximately $N^{1/(\lambda -1)}$ \cite{PhysRevLett.85.4626} and
$k_{\rm min}=2$.

For any network constructed using ANC-DDR with an arbitrary given degree
distribution $P(k)$ (where the distribution is same in both indegree and
outdegree), we can analytically prove (see Appendix \ref{App:AppendixA})
the relationship between the minimal possible directionality $\xi_{\rm
  min}$ and the indegree outdegree correlation $\rho$,
\begin{equation}
\label{equ:31}
\begin{aligned}
E(\xi_{\rm min})&=\frac{1-\rho}{\langle k\rangle}\sum_{k=0}^{N-1}kP(k)\left(\sum_{i=0}^{k}P(i)-\sum_{i=k}^{N-1}P(i)\right).\\
\end{aligned}
\end{equation} 

The simulation and theoretical results concerning the relationship
between $\xi_{\rm min}$ and $\rho$ are consistent, both indicating a
linear relation.  Because of the finite number $M$ of attempts and the
random selection process determining the four-node structure, the
$\xi_{\rm min}$ obtained using simulations is slightly larger than the
theoretical $\xi_{\rm min}$.

Although using ANC-DDR we can construct a network with a given $\rho$
and a given $\xi$ within a corresponding range to $\rho$, it is
computationally expensive to generate a large network with the minimal
possible directionality.  We thus apply DIR to undirected network models
in order to generate directed networks with a directionality ranging
over $[0, 1]$, but with the given indegree and outdegree correlation
$\rho=1$, to understand the effect of directionality on opinion
competitions. We then use ANC-DDR to generate directed networks with a
given indegree and outdegree distribution and correlation, and a given
directionality, to explore the effect of both $\xi$ and $\rho$ on the
opinion model.

\section{The influence of the directionality}

In order to examine how the directionality $\xi$ influences the NCO
model, we apply DIR to undirected network models to generate directed
binomial networks, SF networks, and random regular (RR)
networks\footnote{In this paper, \textit{random regular (RR) networks\/}
  are directed networks in which the indegrees of all nodes and
  outdegrees of all nodes are the same and the nodes are randomly
  connected.} \cite{bollobas2001random} with directionality ranging over
$[0, 1]$. The NCO model is further simulated on each directed network
instance. All simulation results are the average of $10^3$ networks with
$N=10^5$ nodes and $\langle k\rangle=4$.

\begin{figure}
    \centering
    \subfigure[RR]{
      \label{fig:NCO-4_a} 
      \includegraphics[scale=.28]{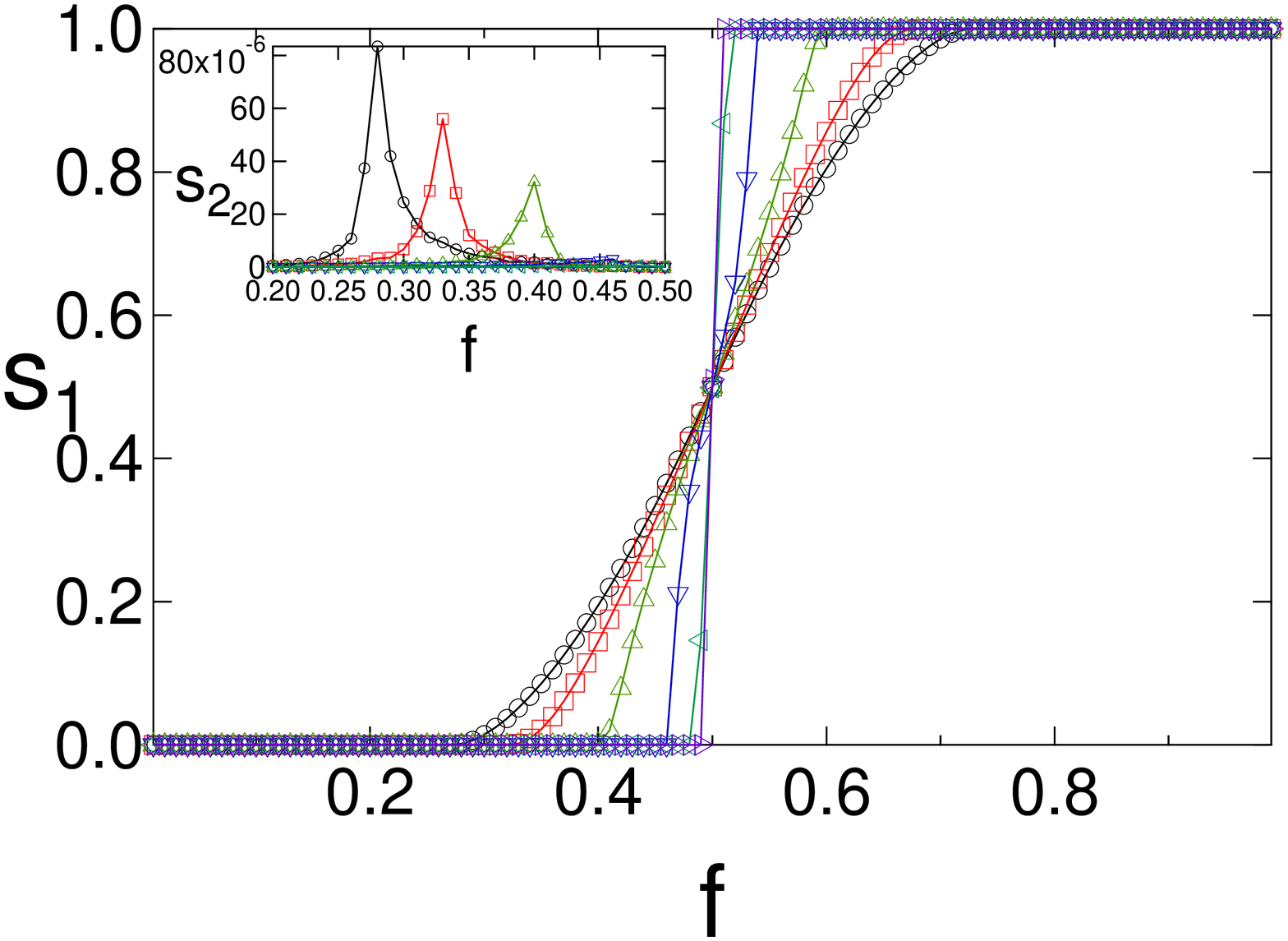}}
    \subfigure[Binomial]{
      \label{fig:NCO-4_b} 
      \includegraphics[scale=.28]{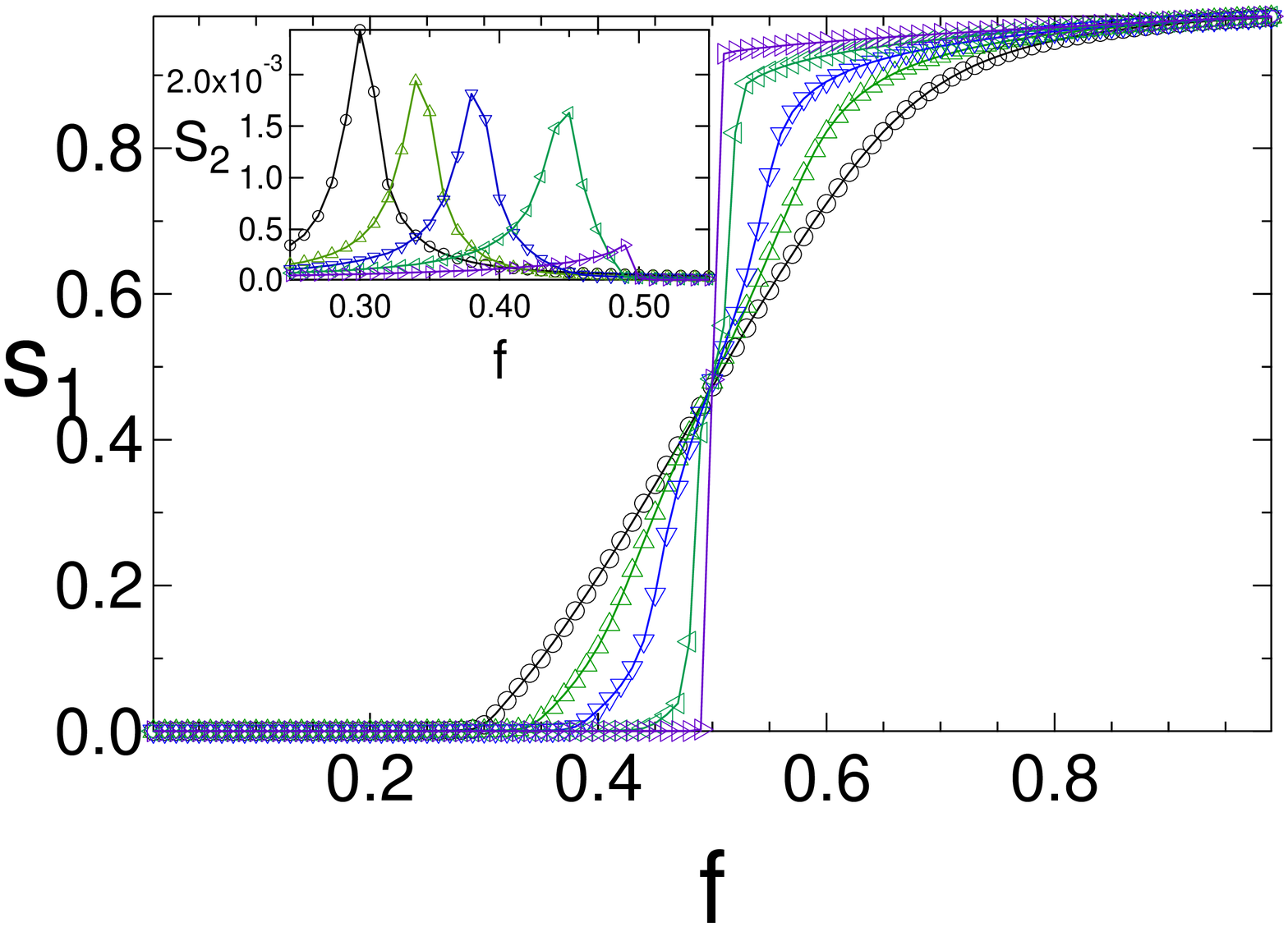}}
    \subfigure[SF]{
      \label{fig:NCO-4_c}
      \includegraphics[scale=.28]{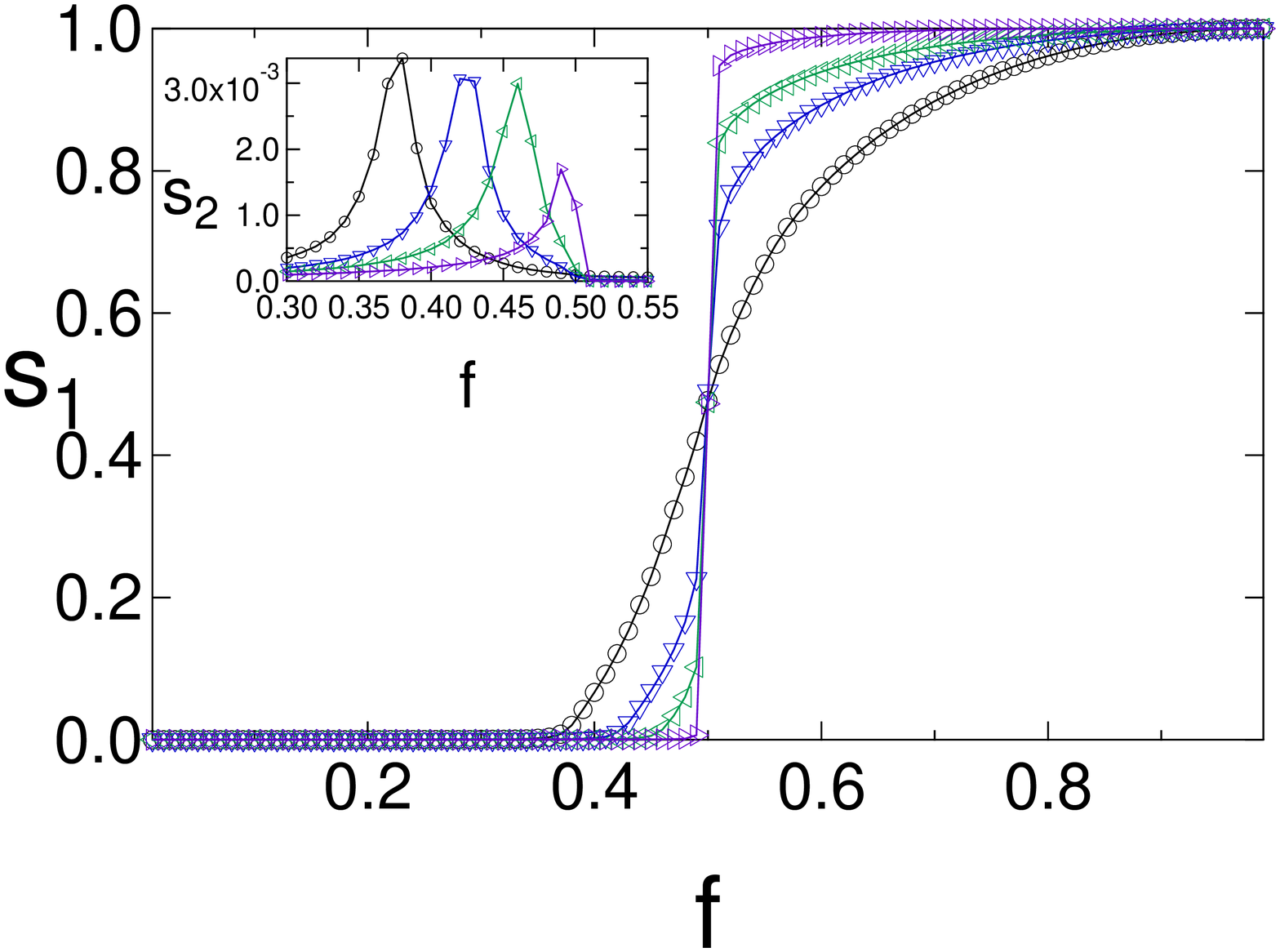}}
    \subfigure[Binomial]{
      \label{fig:NCO-4_d}
      \includegraphics[scale=.28]{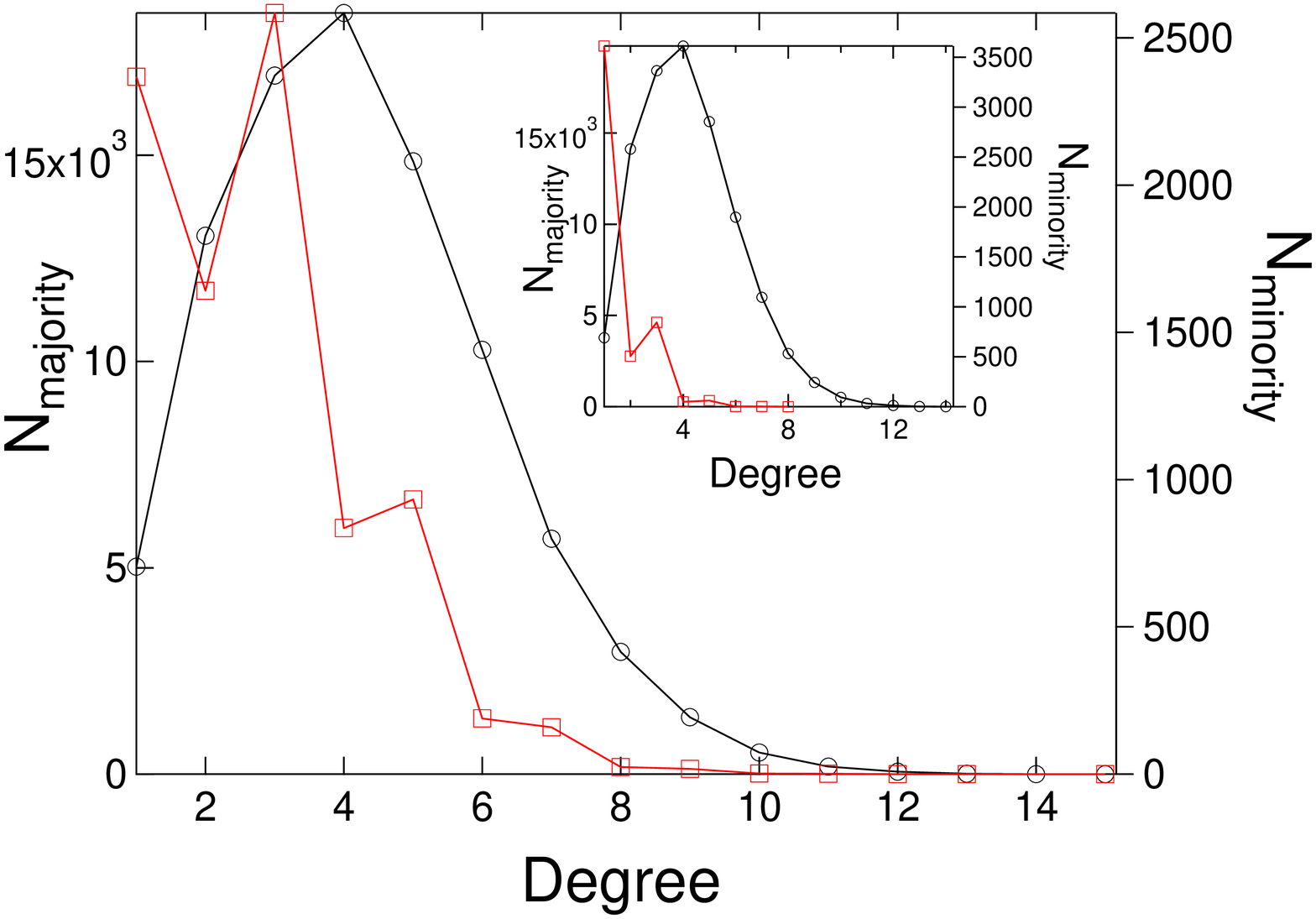}}
    \caption{(Color online) Plot of the normalized largest cluster $s_1$
      of opinion $\sigma_+$ as a function of the initial fraction $f$
      for different values of the directionality $\xi$:
      $0(\textcolor{black}{\circ}), 0.2(\textcolor{red}{\square}),
      0.4(\textcolor{IgorGreen}{\triangle}),
      0.6(\textcolor{IgorBlue}{\triangledown}),
      0.8(\textcolor{IgorGreen1}{\triangleleft}),
      1.0(\textcolor{IgorPurple}{\triangleright})$, and for different
      networks with $N=10^5$ nodes and $\langle k \rangle=4$: (a) RR,
      (b) binomial, (c) SF ($\lambda = 2.63$). In the insets we plot $s_2$ as a function of
      $f$ with the same symbols and for the same networks as in the main
      figure. (d) Plot of the degree distribution of the nodes which
      keep the majority ($\circ$) and the minority opinion
      ($\textcolor{red}{\square}$) in binomial networks (also with $N=10^5$ nodes and $\langle k \rangle=4$), when the
      directionality $\xi=0.0$ (the main figure) and $\xi=1.0$ (the
      inset). All results are based on averaging $1000$ realizations.}
    \label{fig:NCO-4} 
\end{figure}

We use $S_1$ to denote the size of the largest $\sigma_{+}$ cluster in
the steady state (where $\sigma_{+}$ is the initial opinion randomly
assigned to a fraction $f$ of nodes) and $S_2$ to denote the size of the
second largest cluster. For all three types of networks, we plot
$s_1\equiv S_1/N$ and $s_2\equiv S_2/N$ as a function of $f$ for
different values of the directionality $\xi$ in Fig.\ref{fig:NCO-4_a},
\ref{fig:NCO-4_b}, and \ref{fig:NCO-4_c}. Note that, depending on the
value of $\xi$, there is a critical threshold $f\equiv f_c$ above which
there is a giant steady-state component of opinion $\sigma_{+}$. The
peak of $s_2$ indicates the existence of a second-order phase
transition, where $s_1$ is the order parameter and $f$ is the control
parameter. Note that as the value of $\xi$ increases, in all networks
$f_c$ shifts to the right, a shift observable from the shift of the peak
of $s_2$. In RR networks we lose the peak of $s_2$ when the
directionality $\xi$ is close to $1$, which suggests the disappearance of
the second order phase transition. The sharp jump of $s_1$ around
$f=0.5$ also indicates the appearance of an abrupt phase transition.
When these networks contain an increasing one-way influence (increasing
directionality), in all cases the minority opinion will need a greater
number of initial supporters if they are to survive when the steady
state is reached.

To further understand this change we consider two extreme cases, $\xi=0$
and $\xi=1$. In the former, an agent influences only those who can
influence the agent in return. In the latter, an agent influences only
those who cannot influence the agent in return. This latter case allows
a much more rapid spread of opinions, each agent interacts with a larger
number of agents, each has in-neighbors as well as out-neighbors, and
the opinion is diffused over a wider area.  Note that both the majority
and minority opinions can benefit from this wider diffusion, but there
is a higher risk that the minority opinion will be devoured at some
point.  This is the case because the bidirectional link connecting two
minority opinion agents benefits the minority opinion---the two agents
can encourage each other to keep the minority opinion. When rewiring this
kind of link there is a higher probability that the two agents will
interact with the majority opinion and thus a higher probability that
their opinion will be changed to the majority opinion. Thus rewiring
makes it more difficult for the minority opinion to form a stable
structure.

As directionality $\xi$ increases, it is easier for minority opinion
agents to keep their minority opinion if they have fewer
neighbors. Figure~\ref{fig:NCO-4_d} plots the degree distributions (in
which the indegree and outdegree follow the same distribution) of the
minority-opinion nodes and majority-opinion nodes respectively in the
steady state at the critical threshold $f\equiv f_c$ when the
directionality is $\xi=0.0$ and $\xi=1.0$. Note that the degrees of most
of the minority-opinion nodes that keep their minority opinion are equal
to $1$, $2$, or $3$. Minority-opinion nodes with a degree larger than 3
can keep their minority opinion when $\xi=0.0$, but seldom when
$\xi=1.0$--i.e., as the value of $\xi$ increases, the number of nodes
following the majority opinion increases, and only low-degree nodes are
able to keep the minority opinion.

\begin{figure}[htbp]
\centering
\subfigure[]{
\label{fig:NCO-5_a}
\includegraphics[scale=.28]{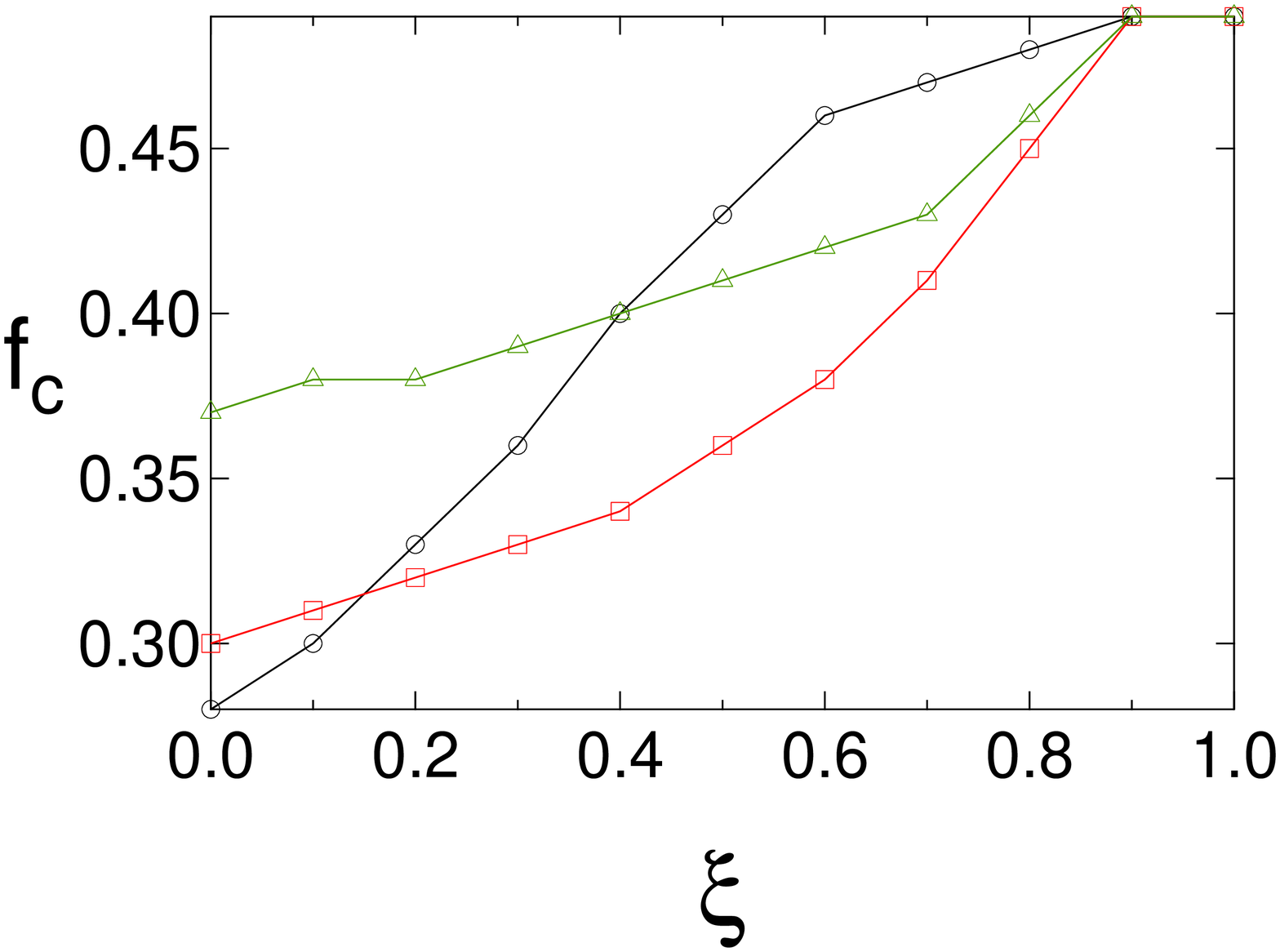}
}
\subfigure[]{
\label{fig:NCO-5_b}
\includegraphics[scale=.28]{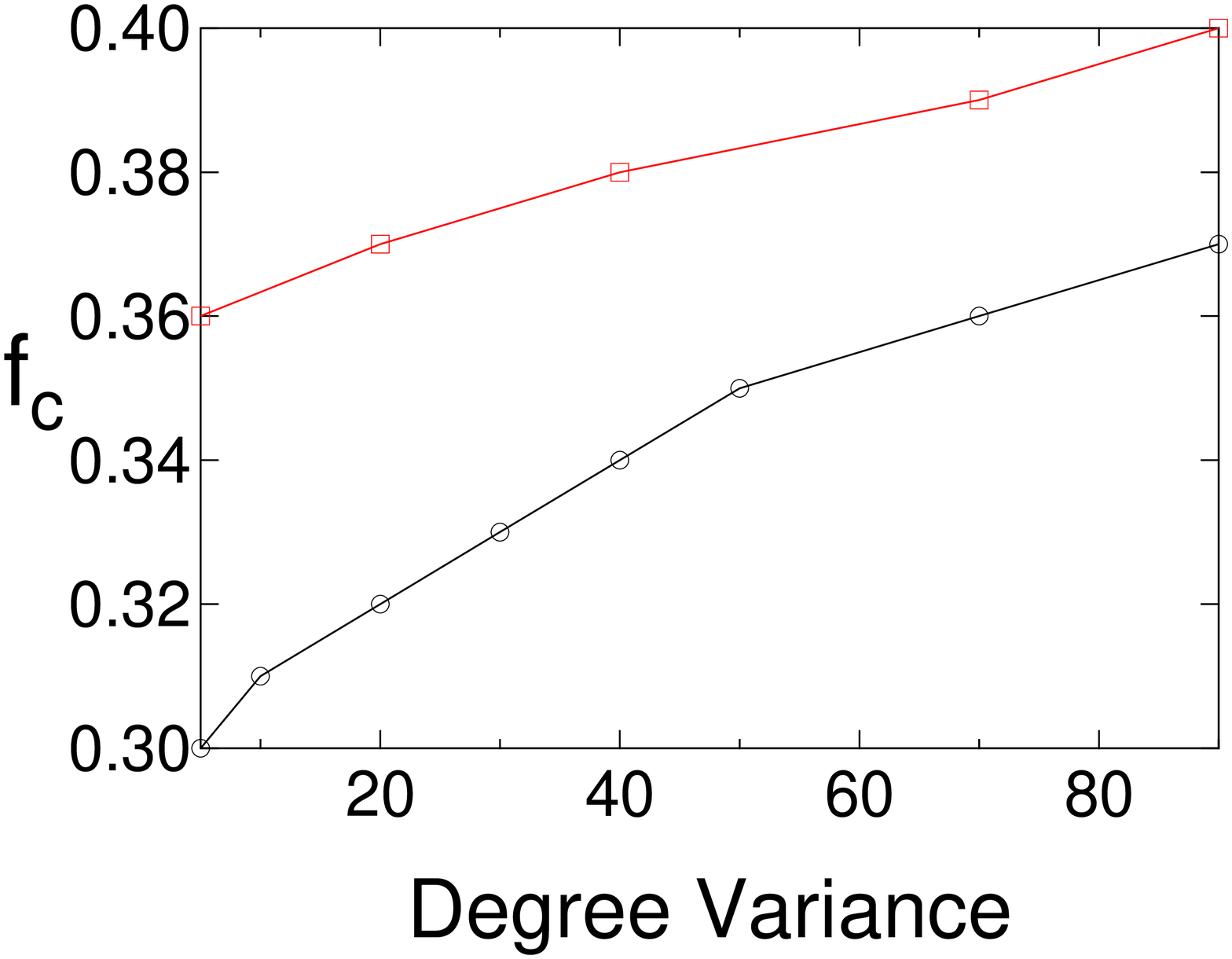}
}
\caption{(Color online) (a) Plot of the critical threshold $f_c$ as a
  function of the directionality $\xi$ for different networks with $N=10^5$ nodes and $\langle k \rangle=4$:
  RR($\circ$), binomial($\textcolor{red}{\square}$) and
  SF($\textcolor{IgorGreen}{\triangle}$) ($\lambda = 2.63$). All results are based on averaging $1000$
    realizations. (b) Plot of the critical
  threshold $f_c$ as a function of the variance of the degree sequence
  of the networks ($N=10^5$ nodes and $\langle k \rangle=4$) with different values of the directionality: $\xi=0$($\circ$) and $\xi=0.5$($\textcolor{red}{\square}$). All results are based on averaging $100$
  realizations.}
\end{figure}

It has been shown that network topology may significantly influence such
dynamic processes in networks as epidemics or cascading failures
\cite{pastor2001epidemic,lopez2005anomalous,motter2004cascade,korniss2007synchronization,boguna2009navigating}.
We thus compare the critical threshold $f_c$ on directed binomial, RR,
and SF networks in which the indegree and outdegree (i) follow the same
binomial distribution, (ii) are a constant, and (iii) follow a power-law
distribution. Figure~\ref{fig:NCO-5_a} shows that, as the directionality
$\xi$ increases, the critical threshold $f_c$ of the RR networks
increases more rapidly than the others. As stated above, as $\xi$
increases, only nodes with degrees less than 4, the average degree, are
likely to keep the minority opinion, and in RR networks all nodal
degrees are $4$. Figure~\ref{fig:NCO-5_a} also shows that the existence
of hubs (extremely high-degree nodes) in SF networks causes them, at $\xi=0$, to have
a much higher critical threshold $f_c$ than the others, and that the
critical threshold in binomial networks is slightly larger than the
critical threshold in RR networks. The existence of hubs benefits the majority-opinion nodes
because the probability that an agent with many friends (i.e., a hub)
will follow the majority opinion and influence many others is high. They
thus strongly contribute to the diffusion of the majority opinion.

Reference \cite{roca2011percolate,sattari2012comment} describes how a
second-order phase transition becomes first-order and the
critical threshold is higher when the average degree increases. In fact,
we find that in the networks with the same average degree, the
larger the variance of the degree sequence, the larger will be its
critical threshold. This is the case because networks with a wider
degree variance are more likely to have majority-opinion hubs that can
influence many other agents. Figure~\ref{fig:NCO-5_b} shows simulation
results that support this behavior. Note that as the variance of the
degree sequence increases, the critical threshold increases. To change
the variance in these simulations we select a SF network with an
average degree $\langle k\rangle=4$, randomly remove an existing link,
and randomly add a link between two nodes previously unconnected.  As we
remove and add links repeatedly, the variance of the degree sequence
decreases and we stop at an excepted variance. To obtain the specified directionality, we apply DIR on the networks. This
gives us a wide range of degree variance, which allows us to study the
relationship between the variance and critical threshold $f_c$.

\section{The influence of indegree and outdegree asymmetry}

We have discussed how the critical threshold $f_c$ increases as the
directionality increases in networks in which the indegree and outdegree
are the same for each node. The number of in-neighbors and 
out-neighbors of nodes in real-world networks often differ, however.  We
mentioned above how a popular singer can influence many people and not
be influenced in return.  The social network of the singer has many more
out-neighbors than in-neighbors. Because this real-world phenomenon is
so ubiquitous, we now examine how different correlations between the
indegree and outdegree affect opinion competition. 
   
\begin{figure}[htbp]
\centering
\subfigure[Binomial]{
\label{fig:NCO-6}
\includegraphics[scale=.28]{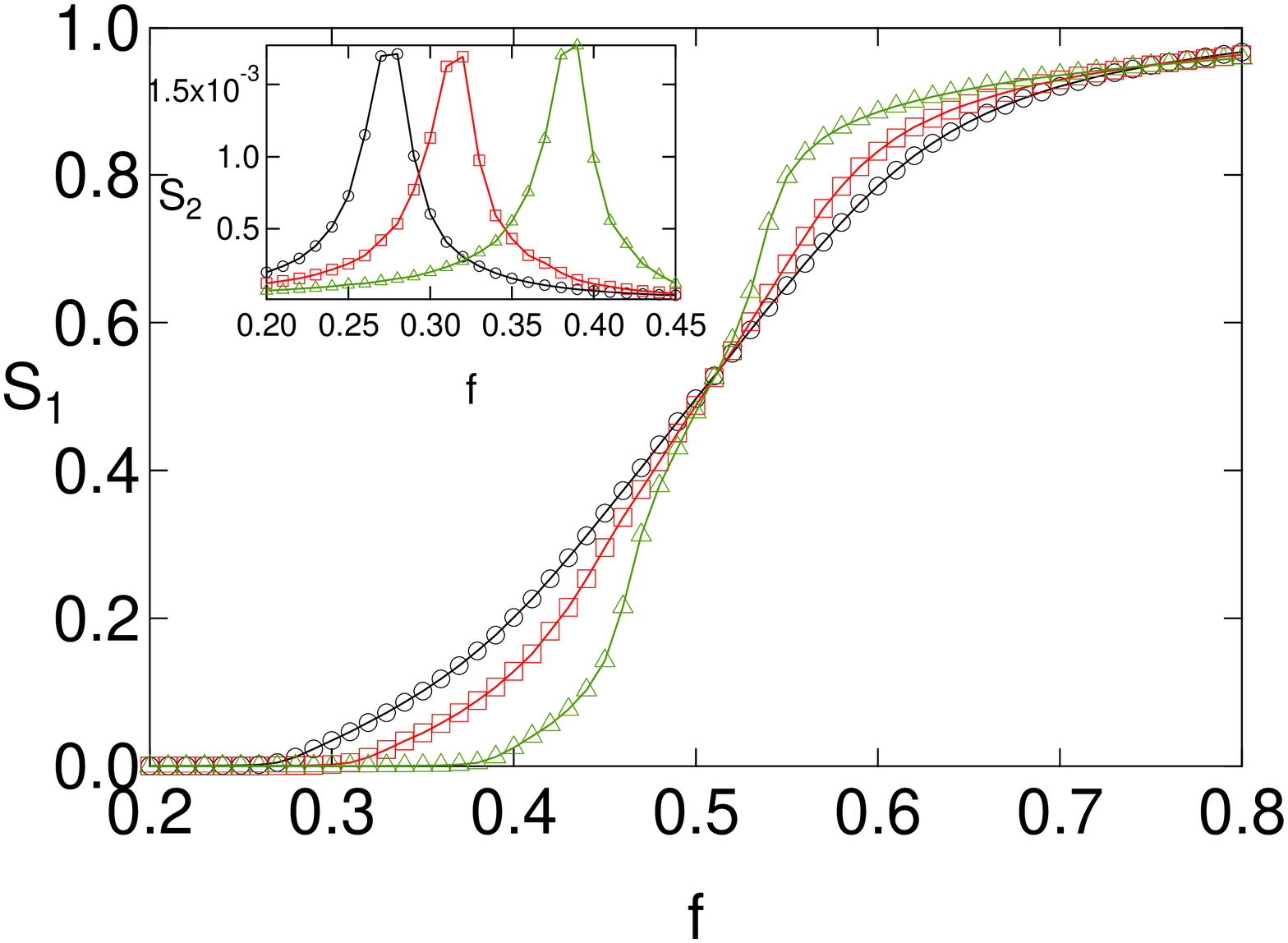}
}
\subfigure[SF]{
\label{fig:NCO-6a}
\includegraphics[scale=.28]{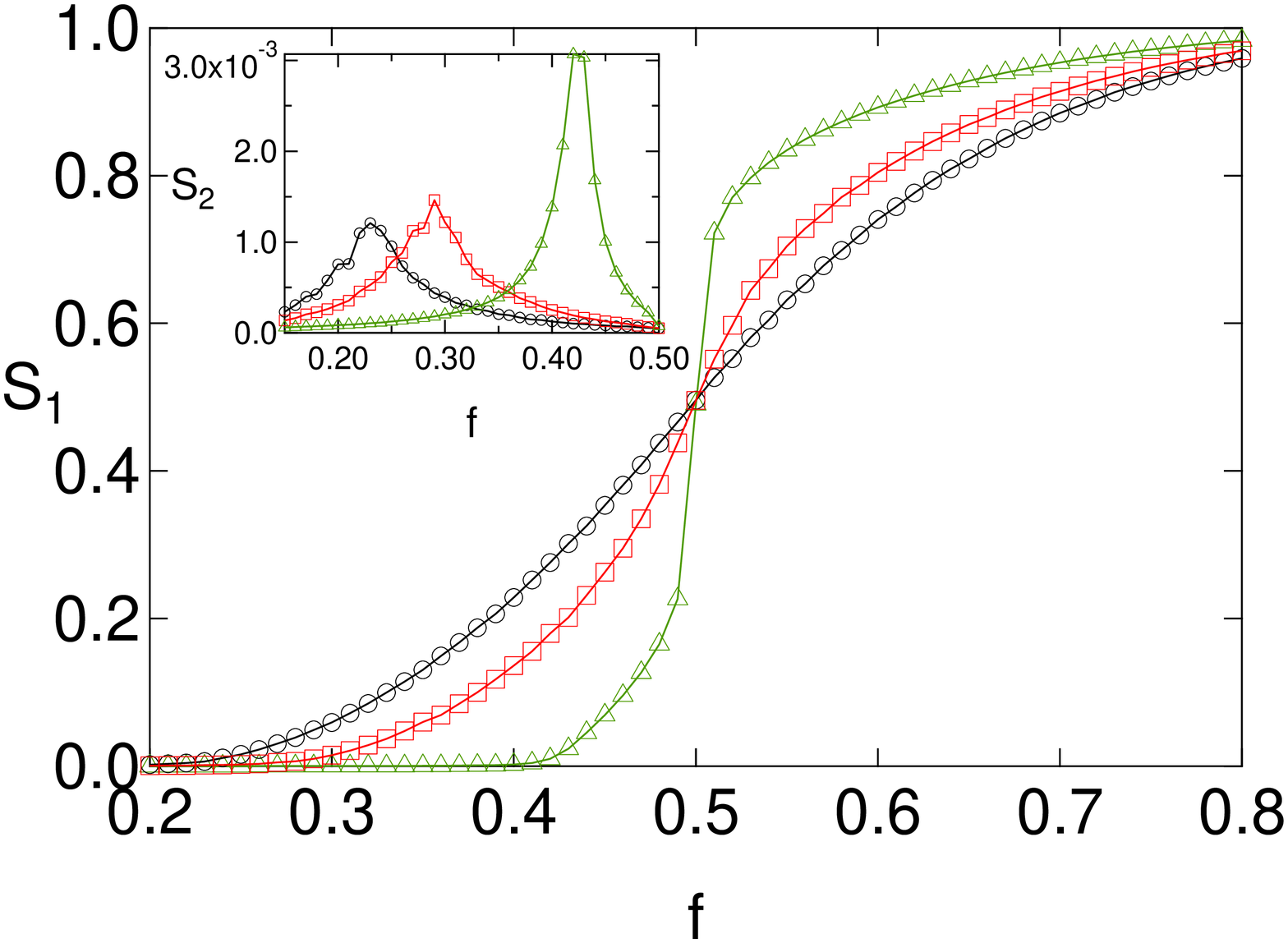}
}
\caption{(Color online) Plot of the normalized largest cluster $s_1$ of
  opinion $\sigma_+$ as a function of the initial fraction $f$, when the
  directionality $\xi=0.6$, for different values of the indegree and
  outdegree correlation $\rho$: $0(\textcolor{black}{\circ}),
  0.5(\textcolor{red}{\square}), 1(\textcolor{IgorGreen}{\triangle}))$,
  and for different networks with $N=10^5$ nodes and $\langle k
  \rangle=4$: (a) Binomial, (b) SF. In the insets we plot $s_2$ as a
  function of $f$ with the same symbols and for the same networks as in
  the main figure. All results are based on averaging $1000$ realizations.}
\end{figure}

\begin{figure}[htbp]
\centering

\includegraphics[scale=.6]{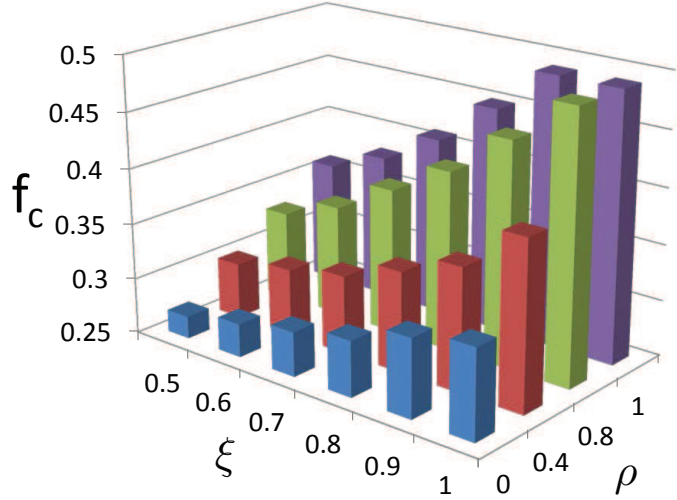}

\caption{(Color online) Plot of the critical threshold $f_c$ as a
  function of the linear correlation coefficient $\rho$ and the
  directionality $\xi$ for binomial networks. All results are based on averaging
  $1000$ realizations.}
  
  \label{fig:NCO-7}
  
\end{figure}

In Section \ref{sec:DecreaseDir} we use ANC-DDR to construct a network
with an arbitrary but identical indegree and outdegree distribution,
together with a given combination of the directionality $\xi$ and the
linear correlation coefficient $\rho$ between the indegree and
outdegree. We perform simulations to study the influence of both the
directionality $\xi$ and the correlation coefficient $\rho$ on the
critical threshold $f_{c}$. Figures~\ref{fig:NCO-6} and \ref{fig:NCO-6a}
show that, given the directionality, the critical threshold increases
for binomial and SF networks, respectively, as the indegree and
outdegree correlation $\rho$ increases.  Figure~\ref{fig:NCO-7} shows
that when the directionality $\xi$ and the correlation coefficient
$\rho$ are increased in binomial networks, the critical threshold
increases. The same behavior is observed in SF networks.

The influence of the indegree and outdegree correlation $\rho$ on the
critical threshold can be understood as follows. A smaller $\rho$ means
a clearer inequality or asymmetry between the indegree and outdegree
links. When the indegree and outdegree links are asymmetrical, a node
with more in-neighbors than out-neighbors is more likely to follow the
majority opinion and, because it has few out-neighbors, its own opinion
will have little influence. Compared with the nodes which have the same number of in-neighbors
and out-neighbors and tends to follow as well as spread the majority opinion, such nodes (with fewer out-neighbors) cannot help. Nodes with more out-neighbors than in-neighbors have greater
influence and can thus hold the minority opinion and contribute to its
spread. Thus the minority opinion benefits more from an inequality
between the indegree and outdegree, or equivalently from a smaller
$\rho$, so the lower correlation coefficient $\rho$ leads to a smaller
critical value $f_c$.

\begin{figure}[htbp]
\centering
\includegraphics[scale=.3]{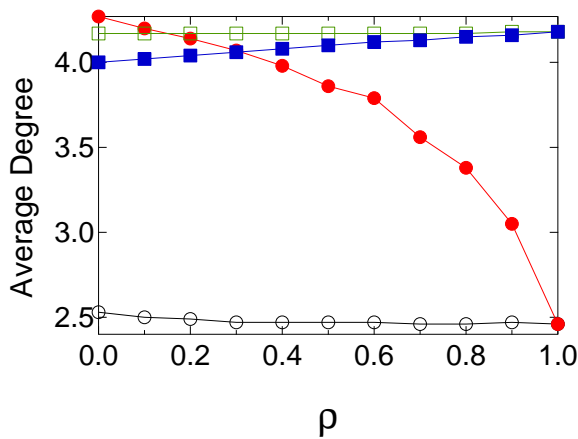}
\caption{(Color online) Plot of the average indegree and outdegree of
  the nodes in the largest $\sigma_+$ and $\sigma_-$ cluster for
  binomial networks, when the initial faction $f$ of the opinion
  $\sigma_+$ equals $0.4$ as a function of $\rho$. The representation of the four lines are as
  follows: the average indegree ($\circ$) and outdegree
  ($\textcolor{red}{\bullet}$) of the nodes in the largest $\sigma_+$
  cluster; the average indegree($\textcolor{IgorGreen}{\square}$) and
  outdegree($\textcolor{blue}{\blacksquare}$) of the nodes in the
  largest $\sigma_-$ cluster. All results are based on averaging $100$
  realizations.}
\label{fig:NCO-8}
\end{figure}

We now further explore the properties of nodes in the final steady
state. We focus on binomial networks in the steady state and calculate as a function of $\rho$,
the average indegree and outdegree in the largest $\sigma_+$ and
$\sigma_-$ clusters with a directionality $\xi=1$ (generated by ANC)
when the initial fraction $f$ of opinion $\sigma_+$ equals $0.4$
(minority). As discussed above, and seen in Fig.~\ref{fig:NCO-8}, the outdegree links of a node with the
minority opinion in the steady state tends to be larger for all $\rho<1$ than
the indegree links, because nodes with few in-neighbors are less
influenced by other nodes and thus can more easily keep their minority
opinion. On the contrary, the indegree of a node with the majority
opinion tends to be larger than its outdegree. Note that, when the initial fraction $f$ of the opinion $\sigma_+$ is
$0.4$, the average number of indegree links is smaller for the nodes in
the largest $\sigma_+$ cluster compared with the nodes in the largest
$\sigma_-$ cluster. Note also that in the majority clusters ($\sigma_+$) both the indegree and the outdegree
are close to $4$, which is the average degree of the whole
network. This is in marked contrast with the average indegree of the nodes in the largest
minority cluster with degree approximately 2.5. The average outdegree of minority is larger
than 4 when the linear correlation coefficient is $\rho=0$. As $\rho$
increases there is a higher correlation between the indegree and
outdegree and the average outdegree of minority decreases rapidly.

\section{Conclusions}

Because of the ubiquity of the non-consensus steady state in real-world
opinion competitions and the dominance of unidirectional relationships
in real-world social networks, we study a non-consensus opinion model on
directed networks. To quantify the extent to which a network is
directed, we use a directionality parameter $\xi$, defined as the ratio
between the number of unidirectional links and the total number of
links.  We also employ a linear correlation coefficient $\rho$ between
the indegree and outdegree to quantify any asymmetry.

We propose two approaches to construct directed networks. The first
is directionality-increasing rewiring (DIR) and is used to rewire the
links of an undirected network to obtain a directed network with any
directionality value $\xi$ without changing the indegree and outdegree,
i.e., the indegree-outdegree correlation value is fixed at $\rho=1$. The
second is ANC-DDR, a combination of asymmetric indegree-outdegree
network construction (ANC) and directionality-decreasing rewiring
(DDR). Using ANC we construct a directed network ($\xi\approx 1$) with
an arbitrary but identical indegree and outdegree distribution and a
given indegree-outdegree correlation $\rho$. We then use DDR to further
decrease the directionality $\xi$ of the network.

We use DIR and ANC-DDR to generate directed networks with a given
combination of $\xi$ and $\rho$ and investigate how the directionality
$\xi$ and the linear correlation coefficient $\rho$ between the indegree
and outdegree links affect the critical threshold $f_c$ of the NCO
model. We find that in both binomial and SF networks increasing $\xi$ or
$\rho$ increases the critical threshold $f_c$. We also find that as $\xi$
and $\rho$ increase, the phase transition becomes abrupt and is no
longer second-order. We find that as a network becomes more directed it
becomes more difficult for a minority opinion to form a cluster, while increasing the indegree-outdegree asymmetry makes the minority
opinion more stable. Our work indicates that
directionality and the asymmetry between indegree and outdegree play a
critical role in real-world opinion competitions.

\section*{Acknowledgements}
\thispagestyle{empty}

We wish to thank ONR (Grants No. N00014-09-1-0380 and No. N00014-12-1-0548), DTRA (Grants No. HDTRA-1-10-1-0014 and No. HDTRA-1-09-1-0035), NSF (Grant No. CMMI 1125290), the European EPIWORK, the Multiplex (No. 317532) EU projects,
CONGAS (Grant No. FP7-ICT-2011-8-317672), BSF, and LINC (No. 289447), the Deutsche Forschungsgemeinschaft (DFG), and the Israel Science Foundation for support.

\newpage
\begin{appendix}
\section{Proof of Eq. (\ref{equ:31})} 
\label{App:AppendixA}

Given the indegree and outdegree of each node, the minimum
directionality can be reached if all the unidirectional links of each
node are either indegree links or outdegree links but not both, because
unidirectional indegree links and outdegree links of a node may form
bidirectional links by rewiring so that the directionality $\xi$ is
further reduced. This means that the minimum directionality can be
reached if an arbitrary node $i$ has only $|K_{i,{\rm in}}-K_{i,{\rm
    out}}|$ unidirectional indegree links or outdegree links but not
both, where $K_{i,{\rm in}}$ and $K_{i,{\rm out}}$ represent the
indegree and outdegree links of node $i$, respectively. Hence the
minimum possible directionality, given the number of indegree and
outdegree links of each node, is
\begin{equation}
\label{Equ:eqA1}
\xi_{\rm min}=\frac{\sum_{i=1}^{N}|K_{i,{\rm in}}-K_{i,{\rm out}}|}{\sum_{i=1}^{N}(K_{i,{\rm in}}+K_{i,{\rm out}})}.     
\end{equation}

We denote the indegree and outdegree sequences by $S_{\rm
  in}=\{K_{i,{\rm in}}|i=1,2,...,N\}$ and $S_{\rm out}=\{K_{i,{\rm
    out}}|i=1,2,...,N\}$ with the same length $N$. The indegree of each
node $K_{i,{\rm in}}$ is independent and follows the distribution $P(k)$
with the mean $\langle k\rangle$. In order to introduce the indegree and
outdegree correlation, ${S_{\rm out}}$ is constructed from ${S_{\rm
    in}}$ as follows: a fraction $\rho$ of the elements in ${S_{\rm
    out}}$ equals that in ${S_{\rm in}}$ ($K_{i,{\rm out}}=K_{i,{\rm
    in}}$, for $i=1,2,...,\rho N$, without loss of generality, we assume
$\rho N$ is an integer), while a fraction $1-\rho$ of ${S_{\rm out}}$ is
obtained by copying and shuffling the rest of ${S_{\rm in}}$, such that
for $i>\rho N$ and large $N$, $K_{i,{\rm in}}$ and $K_{i,{\rm out}}$ are
independent but follow the same distribution $Pr[K=k]=P(k)$. Hence,
\begin{equation}
\begin{aligned}
\label{Equ:eqEXi_min}
E(\xi_{\rm min})&=E(\frac{\sum_{i=\rho N+1}^{N}|K_{i,{\rm in}}-K_{i,{\rm out}}|}{\sum_{i=1}^{N}(K_{i,{\rm in}}+K_{i,{\rm out}})})\\
            &=(1-\rho)E(\frac{N|K_{\rm in}-K_{\rm out}|}{\sum_{i=1}^{N}(K_{i,{\rm in}}+K_{i,{\rm out}})})\\
            &=(1-\rho)E(\xi_{min,\rho=0}),
\end{aligned}
\end{equation} 
where $K_{\rm in}$ and $K_{\rm out}$ are independent random variables
following the same probability distribution $P(k)$, and
$\xi_{min,\rho=0}$ indicates the value of $\xi_{\rm min}$ when $\rho=0$.

We then consider the case when $\rho=0$, i.e.,
\begin{equation}
\begin{aligned}
\label{Equ:eqEXi_minr0}
E(\xi_{{\rm min},\rho=0})&=\frac{E[{\rm Max}(K_{\rm in}, K_{\rm
      out})]-E[{\rm Min}(K_{\rm in}, K_{\rm out})]}{2\langle k\rangle},
\end{aligned}
\end{equation}
where Max$(\cdots)$ and Min$(\cdots)$ are the maximum and minimum
functions, respectively.

The minimum of random variables ${K_{\rm in}, K_{\rm out}}$ has
the distribution,
\begin{equation}
\begin{aligned}
Pr&[{\rm Min}(K_{\rm in}, K_{\rm out})=k]\\
  &=Pr[K_{\rm in}=k]Pr[K_{\rm out}\geq k]+Pr[K_{\rm out}=k]Pr[K_{\rm in}\geq k]\\
  &=2P(k)\sum_{i=k}^{N-1}P(i),
\end{aligned}
\end{equation}
when the two random variables are independent. In the same way, we have
\begin{equation}
Pr[{\rm Max}(K_{\rm in}, K_{\rm out})=k]=2P(k)\sum_{i=1}^{k}P(i)
\end{equation}
Hence,
\begin{equation}
\begin{aligned}
&E(\xi_{{\rm min},\rho=0})=\frac{1}{\langle k\rangle}{\sum_{k=0}^{N-1}kP(k)\left(\sum_{i=0}^{k}P(i)-\sum_{i=k}^{N-1}P(i)\right)}.
\end{aligned}
\end{equation} 
Combining \ref{Equ:eqEXi_minr0} and \ref{Equ:eqEXi_min}, we have
\begin{equation}
\begin{aligned}
E(\xi_{\rm min})&=\frac{1-\rho}{\langle
  k\rangle}\sum_{k=0}^{N-1}kP(k)\left(\sum_{i=0}^{k}P(i)-\sum_{i=k}^{N-1}P(i)\right). 
\end{aligned}
\end{equation}

\end{appendix}
\end{document}